\documentclass[namedreferences]{solarphysics}

\usepackage[hyperref,optionalrh,solaromanenum]{spr-sola-addons} 
\usepackage{graphicx}                    
\usepackage{amssymb}                    
\usepackage{gensymb}
\usepackage{color}                       
\usepackage{breakurl}                         
\usepackage{ctable}
\hypersetup{colorlinks, citecolor=blue, filecolor=black, linkcolor=black, urlcolor=black}

\makeatletter
\newcommand{\rmnum}[1]{\romannumeral #1}
\newcommand{\Rmnum}[1]{\expandafter\@slowromancap\romannumeral #1@}
\makeatother

\newcommand{\ion}[2]{#1\,{\sc \rmnum{#2}}}
\newcommand{\arcsec}{\hbox{$^{\prime\prime}$}}

\newcommand{\kms}{km\,s$^{-1}$}
\newcommand{\vfp}{$v^{\mathrm{f}}_{\mathrm{p}}$}

\newcommand{\vcmeperp}{$v_{\mathrm{CME}_{\perp}}$}
\newcommand{\acmeperp}{$a_{\mathrm{CME}_{\perp}}$}
\newcommand{\fw}{$f(v_{\mathrm{CME}_{\perp}}, t_{\mathrm{cooling}})$}
\newcommand{\dens}{n_{\mathrm{e}}}
\newcommand{\areacme}{$A_{\mathrm{CME}}$}
\newcommand{\fbdens}{$f(B, \dens)$}

\newcommand{\eg}{\emph{e.g.}}
\newcommand{\ie}{\emph{i.e.}}
\newcommand{\cf}{\emph{cf.}}
\newcommand{\rtii}{Type~\Rmnum{2}}
\newcommand{\gtick}{\textcolor{black}{\checkmark}}
\newcommand{\rcross}{\textcolor{red}{$\times$}}
\newcommand{\bfref}[1]{\textcolor{black}{#1}}

\begin{document}

\begin{article}

\begin{opening}

\title{Understanding the Physical Nature of Coronal ``EIT Waves''}

 \author[addressref=aff1,corref,email={david.long@ucl.ac.uk}]{\inits{D.M.}\fnm{D.M.}~\lnm{Long}}
 \author[addressref={aff2,aff3}]{\inits{D.S.}\fnm{D.S.}~\lnm{Bloomfield}}
 \author[addressref=aff4]{\inits{P.F.}\fnm{P.F.}~\lnm{Chen}}
 \author[addressref=aff5]{\inits{C.}\fnm{C.}~\lnm{Downs}}
 \author[addressref=aff2]{\inits{P.T.}\fnm{P.T.}~\lnm{Gallagher}}
 \author[addressref=aff6]{\inits{R.-Y.}\fnm{R.-Y.}~\lnm{Kwon}}
 \author[addressref=aff7]{\inits{K.}\fnm{K.}~\lnm{Vanninathan}}
 \author[addressref=aff7]{\inits{A.M.}\fnm{A.M.}~\lnm{Veronig}}
 \author[addressref=aff8]{\inits{A.}\fnm{A.}~\lnm{Vourlidas}}
 \author[addressref=aff9]{\inits{B.}\fnm{B.}~\lnm{Vr\v{s}nak}}
 \author[addressref=aff10]{\inits{A.}\fnm{A.}~\lnm{Warmuth}}
 \author[addressref=aff9]{\inits{T.}\fnm{T.}~\lnm{\v{Z}ic}}

\runningauthor{D.M.~Long {\it et al.}}
\runningtitle{Understanding the Physical Nature of ``EIT Waves''}

\address[id=aff1]{UCL-Mullard Space Science Laboratory, Holmbury St.~Mary, Dorking, Surrey, RH5~6NT, UK}
\address[id=aff2]{School of Physics, Trinity College Dublin, College Green, Dublin 2, Ireland}
\address[id=aff3]{Northumbria University, Newcastle upon Tyne, NE1~8ST, UK}
\address[id=aff4]{School of Astronomy \& Space Science, Nanjing University, 163 Xianlin Ave, Nanjing 210023, China PR}
\address[id=aff5]{Predictive Science Inc., 9990 Mesa Rim Rd, Suite 170, San Diego, CA~92121, USA}
\address[id=aff6]{College of Science, George Mason University, 4400 University Drive, Fairfax, VA~22030, USA}
\address[id=aff7]{Kanzelh\"{o}he Observatory/IGAM, Institute of Physics, University of Graz, 8010 Graz, Austria}
\address[id=aff8]{The Johns Hopkins University Applied Physics Laboratory, Laurel, MD~20723, USA}
\address[id=aff9]{Hvar Observatory, Faculty of Geodesy, Kaciceva 26, 10000 Zagreb, Croatia}
\address[id=aff10]{Leibniz-Institut f\"{u}r Astrophysik Potsdam (AIP), An der Sternwarte 16, 14482 Potsdam, Germany}

\begin{abstract}
For almost 20 years the physical nature of globally propagating waves in the solar corona (commonly called 
``EIT waves'') has been controversial and subject to debate. Additional theories have been proposed over 
the years to explain observations that did not fit with the originally proposed fast-mode wave interpretation. 
However, the incompatibility of observations made using the \emph{Extreme-ultraviolet Imaging Telescope} (EIT) onboard the 
\emph{Solar and Heliospheric Observatory} with the fast-mode wave interpretation was challenged by 
differing viewpoints from the twin \emph{Solar Terrestrial Relations Observatory} spacecraft and higher 
spatial/temporal resolution data from the \emph{Solar Dynamics Observatory}. In this article, we reexamine the 
theories proposed to explain ``EIT waves'' to identify measurable properties and behaviours that can be 
compared to current and future observations. Most of us conclude that ``EIT waves'' are best described as 
fast-mode large-amplitude waves/shocks that are initially driven by the impulsive expansion of an erupting 
coronal mass ejection in the low corona.
\end{abstract}

\keywords{Coronal Mass Ejections, Low Coronal Signatures; Waves, Magnetohydrodynamic; Waves, Propagation; 
Waves, Shock}

\end{opening}

\section{Introduction}\label{s:intro} 

Globally propagating waves in the solar corona have been studied in detail since being first directly 
observed by the \emph{Extreme-ultraviolet Imaging Telescope} \citep[EIT:][]{Dela:1995} onboard the \emph{Solar and 
Heliospheric Observatory} \citep[SOHO:][]{Domingo:1995}. However, a physical explanation for ``EIT waves'' 
(as they are commonly called) has remained elusive due to a paucity of observations and inconsistent analyses. 
This has led to the continued development of competing theories designed to explain the phenomenon. 

``EIT waves'' are generally observed as bright pulses in the low solar corona emanating from the source of a 
solar eruption, and often traverse the solar disk in less than an hour. They can have velocities of up to 
$\approx$\,1400\,\kms\ \citep[\cf][]{Nitta:2013}, but are most typically observed at velocities of 
200\,--\,500\,\kms\ \citep{Klassen:2000,Thompson:2009,Muhr:2014}. It was initially suggested that they were 
magnetohydrodynamic (MHD) fast-mode waves 
 driven either by the erupting coronal mass ejection (CME) or alternatively by the 
associated flare \citep[\eg][]{Moses:1997,Dere:1997,Thompson:1998}. This was consistent with the global MHD 
fast-mode wave propagating in the corona that was predicted by \citet{Uchida:1968} to explain the chromospheric 
Moreton--Ramsey wave \citep{Moreton:1960a,Moreton:1960b}. 

However, 
 ``EIT wave'' velocities were found to be much lower than 
 estimated \bfref{quiet-Sun} coronal 
 \bfref{fast-mode speeds, leading some to} 
 suggest that they could not be 
 fast-mode waves. 
 While ``EIT waves'' exhibit \bfref{the wave attributes of}
 reflection \citep[\eg][]{Gopal:2009}, refraction 
\citep[\eg][]{Wills-Davey:1999} and \bfref{transmission} \citep[\eg][]{Olmedo:2012}, they can 
 remain stationary at coronal hole (CH) boundaries for \bfref{tens of minutes to hours \citep[\eg][]{Delannee:2000}} -- behaviour 
 \bfref{originally proposed as} 
 inconsistent with the wave interpretation.

These discrepancies led to the development of several alternative explanations for ``EIT waves''. One branch 
elaborated on the wave interpretation, treating ``EIT waves'' as slow-mode waves \citep[\cf][]{Wang:2009}, 
 slow-mode solitons \citep[\eg][]{Wills-Davey:2007} or more generally as shock waves \citep[or large-amplitude 
MHD waves, for more details see the review by][]{Vrsnak:2008}. The other branch eschewed waves 
 entirely, instead treating 
 them 
 as \bfref{pseudo-waves} resulting from coronal magnetic field reconfiguration during 
 CME eruption. In this approach, 
 ``EIT wave'' 
 brightenings 
 result from several 
 different 
 processes, including stretching of magnetic-field lines \citep[\cf][]{Chen:2002}, Joule heating in a 
 current shell \citep[\cf][]{Delannee:2007} or continuous small-scale reconnection \citep[\cf][]{Attrill:2007}.

As with other aspects of solar eruptive events, ``EIT waves'' are a relatively common phenomenon. Although 
they are 
 less common than CMEs \citep[every ``EIT wave'' has an associated CME, but the converse is not 
necessarily true;][]{Biesecker:2002}, between 1997 and 2013 at least 407 events have been identified using the 
SOHO, \emph{Solar Terrestrial Relations Observatory} \citep[STEREO:][]{Kaiser:2008} and \emph{Solar Dynamics 
Observatory} \citep[SDO:][]{Pesnell:2012} spacecraft \citep[\cf][]{Thompson:2009,Nitta:2013,Muhr:2014}. Despite 
this, ``EIT waves'' tend to be studied in isolation, using single-event studies to make generalised statements 
about their physical interpretation. This approach 
 led to 
 a disconnect 
between advocates of the wave and pseudo-wave interpretations, with both sides using different (and in most 
cases single-event) observations to support their preferred view. 

The majority of the theories designed to explain this phenomenon were originally proposed based on observations
from SOHO/EIT that had a spatial and temporal sampling of $\approx$\,5\arcsec\ and \bfref{12\,--}\,15\,minutes, respectively.
This typically provided two observations of an ``EIT wave'' per event, and it places restrictions on the resulting 
physical interpretation. However, this was improved on by the \emph{Extreme UltraViolet Imager} 
\citep[EUVI:][]{Wuelser:2004} onboard the twin STEREO spacecraft ($\approx$\,3.2\arcsec\ and \bfref{1.25}\,--\,10\,minutes) 
and more recently the \emph{Atmospheric Imaging Assembly} \citep[AIA:][]{Lemen:2012} onboard SDO ($\approx$\,1.2\arcsec\ 
and 12\,seconds). Although this improvement in both spatial and temporal resolution should allow a more 
rigorous testing of all of the different interpretations for ``EIT waves'', this has not been the case, primarily 
because very few testable predictions of physical properties and behaviour are provided for each proposed theory.

This article should be viewed as being complementary to the reviews of ``EIT waves'' by \citet{Wills-davey:2009}, 
\citet{Gallagher:2011}, \citet{Zhukov:2011}, \citet{Patsourakos:2012}, \citet{Liu:2014}, and in particular the 
recent reviews by \citet{Warmuth:2015} \bfref{and \citet{Chen:2016}. This is achieved by identifying all of the 
currently measurable properties (and some beyond our current capabilities) for each of the theories/models. A 
direct 
 comparison is then performed for each of these properties, making use of the most recently published 
results (\eg\ observations of differential emission measure and simulations that relate ``EIT waves'' to other 
solar phenomena). This effort is a result of} an International Working Team on ``The Nature of Coronal Bright 
Fronts'' convened at the International Space Science Institute (ISSI: \url{http://www.issibern.ch}).

In this article, we aim to identify and quantify the physical properties and behaviour predicted by the theories
proposed to explain the ``EIT wave'' phenomenon, \bfref{greatly expanding on the initial attempt by \citet{Patsourakos:2009}}. 
Each theory is outlined in Section~\ref{s:theory} with particular 
emphasis placed on what they each predict for a variety of physical properties, including kinematics, height, 
bounded area, and variation in density, temperature and magnetic-field strength. The analysis techniques currently 
used to identify and study ``EIT waves'' and the limitations that they naturally impose on the observations are 
outlined in Section~\ref{s:analysis}, allowing the optimal technique for each property to be identified. Finally, 
the best interpretation for ``EIT waves'' given current analysis techniques is summarised in Section~\ref{s:disc}.

\section{Theories}\label{s:theory}

\ctable[
caption = {Prediction of physical properties of pulses from theory},
label = tbl:theories,
center,
sideways
]{lcccccc}{
\tnote[$^\mathrm{a}$]{describing only the \bfref{slower component of the two-wave scenario (\ie\ the density 
 perturbation component)}}
\tnote[$^\mathrm{b}$]{\bfref{$U=I_{\mathrm{peak}}/I_{0}$ (\ie\ the ratio of peak intensity [$I_{\mathrm{peak}}$] 
 to background intensity [$I_{0}$])}}
\tnote[$^\mathrm{c}$]{\citet{Delannee:2008}}
\tnote[$^\mathrm{d}$]{height of adjacent small-scale loops; value quoted \bfref{in} \citet{Patsourakos:2009b}}
}{
\FL
Pulse                       & \multicolumn{3}{c}{Wave theories}                            & \multicolumn{3}{c}{Pseudo-wave theories/models}\NN
physical                    & \multicolumn{2}{c}{Fast-mode}               & Slow-mode      & Field-line                      & Current                                    & Continuous\NN
property                    & Small amp.     & Large amp.                 & soliton        & stretching\tmark[$^\mathrm{a}$] & shell                                      & reconnection\NN
                            & linear wave    & wave/shock                 &                &                                 &                                            & 
\ML
Phase velocity, $v$  & \vfp           & $>$\,\vfp ; $\propto$\,$U$\tmark[$^\mathrm{b}$] & $\propto$\,$U$\tmark[$^\mathrm{b}$] & $<$\,\vfp                      & \vcmeperp                                   & \vcmeperp\NN
                     & $>$\,\vcmeperp & $>$\vcmeperp               & \ldots         & \vcmeperp                       & \ldots                                      & \ldots\NN
Acceleration, $a$    & 0              & $<$\,0                     & 0              & \acmeperp                       & \acmeperp                                   & \acmeperp\NN
Broadening           & $\sim$\,0      & $>$\,0                     & 0              & $>$\,0                          & $f$(\acmeperp)                              & \fw\NN
$\Delta B$           & $\gtrsim$\,0   & $>$\,0                     & $<$\,0         & $>$\,0                          & $\approx$\,0                                & $<$\,0\NN
$\Delta T$           & Adia.          & Adia. + $Q$                & Adia.          & Adia.                           & $Q_{\mathrm{Joule}}$ + adia.                & Non-adia.\NN
$\Delta \dens$       & Compression    & Compression                & Compression    & Compression                     & Compression                                 & Upflows\NN
Height               & \fbdens        & \fbdens                    & \fbdens        & $f$(CME bubble)                 & $\approx$\,280 or 407\,Mm\tmark[$^\mathrm{c}$] & $<$\,10\,Mm\tmark[$^\mathrm{d}$]\NN
Area bounded         & $>$\,\areacme  & $>$\,\areacme              & $>$\,\areacme  & \areacme                        & \areacme                                    & \areacme\NN
Rotation             & Possible       & Possible                   & Possible       & Possible                        & Possible                                    & Possible\NN
Reflection           & Yes            & Yes                        & Possible       & No                              & No                                          & No\NN
Refraction           & Yes            & Yes                        & Possible       & No                              & No                                          & No\NN
\bfref{Transmission} & Yes            & Yes                        & Yes            & No                              & No                                          & No\NN
Stationary fronts    & \bfref{Yes}            & \bfref{Yes}                        & No             & Yes                             & Yes                                         & Yes\NN
Co-spatial \rtii     & No             & Possible                   & No             & No                              & No                                          & No\NN
Moreton wave         & No             & Possible                   & No             & No                              & No                                          & Possible
\LL
}

The development of multiple theories to explain ``EIT waves'' can be primarily attributed to inconsistencies in 
interpretation. This \bfref{was} compounded by the small number of events studied in detail by multiple authors, 
\bfref{including (but not limited to)} the events from 12 May 1997 \citep[\eg][]{Moses:1997,Dere:1997,Thompson:1998}, 
19 May 2007 \citep[\eg][]{Long:2008,Veronig:2008,Gopal:2009,Attrill:2010}, 13 February 2009 
\citep[\eg][]{Patsourakos:2009a,Cohen:2009,Kienreich:2009}, and 15 February 2011 
\citep[\eg][]{Schrijver:2011,Olmedo:2012,Vanninathan:2015}. This approach \bfref{(initially driven by the relatively 
small number of well-observed events in the SOHO/EIT era)} led to a situation where the theories proposed to explain 
``EIT waves'' were \bfref{primarily} developed to explain the behaviour of individual and necessarily different events, 
while paying minimal attention to predicting more generalised behaviour and observables that may help to understand 
their true nature. \bfref{The launches of 
 STEREO and SDO have led to more statistical ``EIT wave'' studies using the analysis techniques 
 developed for individual events},
but these have focused on individual properties such as kinematics \bfref{and wave-pulse characteristics} 
\citep[\eg][]{Thompson:2009,Warmuth:2011,Nitta:2013,Muhr:2014}.

There are two main branches of proposed theories -- wave (Table~\ref{tbl:theories}, columns 2\,--\,4) and 
pseudo-wave (Table~\ref{tbl:theories}, columns 5\,--\,7). In the wave interpretation, ``EIT waves'' are 
classified using the MHD wave equations as linear fast-mode waves \citep[\eg][]{Thompson:1998}, or 
alternatively as nonlinear waves such as large-amplitude fast-mode or shock waves \citep[\eg][]{Vrsnak:2008} or MHD 
slow-mode solitons \citep[\cf][]{Wills-Davey:2007}. In the pseudo-wave interpretation they are described as 
brightenings arising from magnetic-field-line stretching \citep{Chen:2002}, Joule heating in current shells 
\citep{Delannee:2007,Delannee:2008} or continuous small-scale reconnection \citep{Attrill:2007}. 
The physical processes demanded by these different interpretations should result in different observed 
behaviour, thus providing an opportunity to distinguish between theories. 

However, a detailed discussion of 
 properties and behaviour predicted by each theory was often omitted in 
their initial presentation, with the result that conclusions about their validity continue to be made based on 
observations of single properties \bfref{(\eg\ kinematics or pulse characteristics)}. In this section, we attempt to overcome this 
issue and identify 
 observable properties predicted by each of the different theories and models proposed 
to explain the ``EIT wave'' phenomenon. The predicted properties for each are presented in 
Table~\ref{tbl:theories} under the assumption of an idealised homogeneous background corona, while 
Sections~\ref{ss:mhdwave} to \ref{ss:attrill} outline the individual theories and the reasoning behind their 
physical predictions.

\subsection{MHD Fast-Mode Waves}\label{ss:mhdwave}

\begin{figure*}[!t]
\centering{
	\includegraphics[width=0.99\textwidth,clip=,trim=0mm 0mm 0mm 0mm]{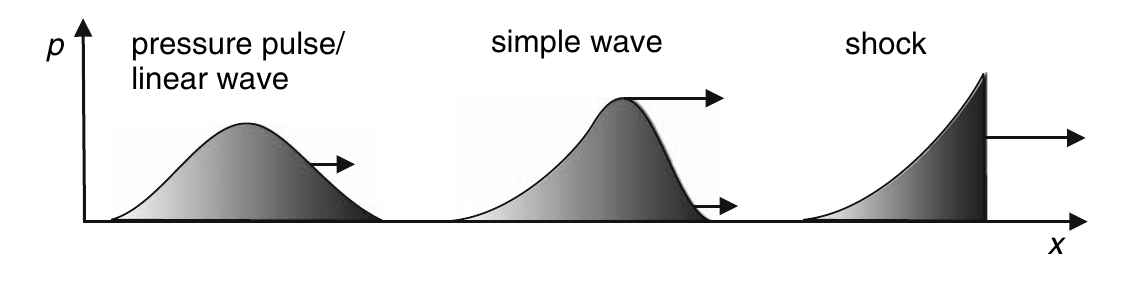}
    }
\caption{\bfref{Graphical representation of the wave models, highlighting the differences between pulse wave forms 
		\citep[adapted from Figure~2 of][]{Warmuth:2007}. The linear slow-mode and fast-mode waves involve small perturbations 
        and so take the pulse form in the left panel. Nonlinear effects become more important as the amplitude 
        increases, with the pulse taking the form of a simple wave in the centre panel. A special solution of the 
        nonlinear wave equations can involve this steepening being canceled out by dispersive effects, leading to 
        the formation of a MHD soliton. Alternatively, the simple wave may become shocked, taking the pulse form 
        in the right panel.}}
\label{fig:wave}
\end{figure*}

The MHD wave equations may be solved to produce two linearised magnetoacoustic wave types \bfref{(with 
 forms depicted in Figure~\ref{fig:wave}, left panel)}, namely slow- 
 and fast-mode waves \citep[\eg][]{Priest:2014}. 
 For 
 both, 
 the 
 phase velocity
($v_{\mathrm{p}}^{\mathrm{s}}$ and \vfp, respectively) is defined by the sound speed [$c_{\mathrm{s}}$] and the
Alfv\'{e}n velocity [$v_{\mathrm{A}}$] of the medium through which they propagate (\ie\ the 
 corona), but 
also by the angle between the direction of propagation and 
 that 
 of the magnetic field. Slow-mode 
waves have velocities of 0\,\kms\ perpendicular to the magnetic field, while for fast-mode waves it is  
dependent on the temperature, density, and magnetic field 
 in 
 the corona. ``EIT waves'' 
 travel across the Sun where the 
 field is primarily radial, so they cannot 
be interpreted as slow-mode waves. 

In addition to linear forms of the MHD fast-mode wave, ``EIT waves'' can also be interpreted as large-amplitude 
pulses using nonlinear wave theory as discussed by \citet{Warmuth:2007}, \citet{Vrsnak:2008}, and \citet{Lulic:2013}. 
With this approach, if the amplitude of the wave is sufficiently large the nonlinear terms become important and the 
crest of the wave can move faster than the characteristic speed of the medium through which it is passing. This 
so-called simple wave \citep[\cf][]{Mann:1995a} therefore begins to steepen \bfref{(Figure~\ref{fig:wave}, centre panel)}, and may ultimately form a shock wave \bfref{(Figure~\ref{fig:wave}, right panel)}. 
This interpretation of ``EIT waves'' as global shock waves was initially motivated by their strong association with 
metric \rtii\ radio bursts that indicate the presence of a shock front 
\citep[\eg][]{Klassen:2000,Biesecker:2002,Vrsnak:2008}.

The most simple definition of a shock wave is of a discontinuity travelling faster than the characteristic 
speed of the ambient medium through which it is propagating; in this case the fast-mode velocity of the solar 
corona. However, it is possible for a piston-driven shock (wherein the motion of a piston drives a shock that the 
medium cannot flow around) to be formed even if the piston has a velocity lower than the characteristic speed of 
the medium. In this situation, it is also possible for the shock to travel faster than and therefore decouple from 
the piston before propagating freely. Interpretation as a shock front raised discussion as to the initial driver, 
with opposing camps claiming that the ``EIT wave'' was initially driven either by the rapid release of energy 
during the impulsive phase of a flare \citep[\cf][]{Vrsnak:2001} or alternatively by the rapid expansion of a CME 
in the low corona \citep[\eg][]{Cliver:2005}. However, both camps agreed that, once formed, the ``EIT wave'' 
would then decouple from its driver and become a freely propagating shocked simple wave \citep[\cf][]{Landau:1959} 
that decelerates, consistent with the piston-driven shock interpretation. 

As well as the velocity and acceleration of the pulse, variations in pulse width can be compared to that predicted 
by theory. Both of the fast-mode wave types introduced above exhibit 
 broadening during propagation. 
For both small- and large-amplitude waves this may result from 
 superposition of multiple frequencies within 
the pulse (\ie\ dispersion), while 
 wave 
 steepening may also contribute to broadening for large-amplitude 
waves/shocks. In terms of physical properties, an increase in magnetic-field strength [$B$], should be produced 
 during 
the passage of both fast-mode 
 types \citep[][]{Priest:1982}, but 
 it 
 will be negligible 
 for 
 small-amplitude waves. 
 Small- and large-amplitude waves both 
 produce temperature and 
 density increases, which should be non-adiabatic for shocks and adiabatic otherwise.

Geometrically, the vertical extent of both wave types should be dependent on how the density and magnetic field 
vary with height above the Sun. As both are initially driven by the erupting CME, their spatial extent at any 
point in time should match or exceed the spatial extent of the associated CME. The wave front could appear to 
rotate during the driven phase if the CME driver is elliptical and itself rotates. As these are true wave 
solutions, they are expected to behave as such, undergoing reflection and refraction where appropriate. A direct 
consequence of this is the transmission of a portion of the wave front through a CH and apparent stationary wave fronts 
at the CH boundary.

In terms of additional phenomena associated with fast-mode waves, only the large-amplitude wave/shock provides 
the necessary conditions to produce co-spatial \rtii\ radio bursts (through the presence of a shock) and 
Moreton--Ramsey waves (given sufficient pressure acting downwards on the chromosphere).

\subsection{MHD Slow-Mode Solitons}\label{ss:soliton}

The concept of the ``EIT wave'' as an MHD slow-mode soliton was proposed by \citet{Wills-Davey:2007} in an 
attempt to explain some of the discrepancies between the observed properties of ``EIT waves'' and predictions 
of linear MHD fast-mode wave theory. In particular, the authors identified several issues where predictions did not 
match observations, namely the value and variety of observed pulse velocities and what this means for the theoretical 
assumption of a low-$\beta$ plasma in the corona and the coherence of the pulse over the duration of its 
observation. It was argued that these issues made the MHD fast-mode wave interpretation unfeasible, instead 
suggesting that they were most consistent with the interpretation of the pulse as an MHD slow-mode soliton. 

Although \citet{Wills-Davey:2007} only give a brief qualitative argument for solitons as a candidate mechanism for 
``EIT waves'', we can estimate some of their physical properties here. The MHD slow-mode soliton provides a special 
solution to the nonlinear MHD wave equations, with the nonlinear steepening of the wave 
 \bfref{(Figure~\ref{fig:wave}, centre panel)} exactly canceled out 
by dispersive effects. This allows a wave packet (\ie\ 
 soliton) to form that is observed as a bright pulse 
propagating at constant velocity (\ie\ $a=0$) and width (\ie\ no broadening). The 
 velocity is dependent on the amplitude of the pulse intensity, $U=I_{\mathrm{peak}}/I_{0}$ \bfref{(\ie\ the ratio 
of peak intensity [$I_{\mathrm{peak}}$] to background intensity [$I_{0}$)}] indicating that 
brighter, 
 higher-amplitude pulses will exhibit greater velocities.

As with the fast-mode waves in Section~\ref{ss:mhdwave}, the MHD slow-mode soliton should result in an adiabatic 
increase in both temperature and density, although this is accompanied by a decrease in magnetic-field strength 
(since the gas and magnetic pressures are out of phase for slow-mode waves). Similarly, the vertical extent of 
the soliton will be a function of the background magnetic-field strength and density, while the lateral extent will 
exceed that of the associated CME. In addition, an MHD slow-mode soliton should exhibit rotation with propagation 
given a rotating elliptical driver, and may undergo reflection and refraction under specific conditions. MHD slow-mode solitons do not interact with CH boundaries in the same manner as fast-mode waves/shocks, resulting in their 
transmission through CHs without producing stationary fronts.

The MHD slow-mode soliton is not a shock and so does not produce the necessary conditions for either a co-spatial 
\rtii\ radio burst or a chromospheric Moreton-Ramsey wave, similar to the small-amplitude fast-mode wave.

\subsection{Field-Line Stretching Model}\label{ss:2wave}

\begin{figure*}[!t]
\centering{
	\includegraphics[width=0.95\textwidth,clip=,trim=0mm 0mm 0mm 0mm]{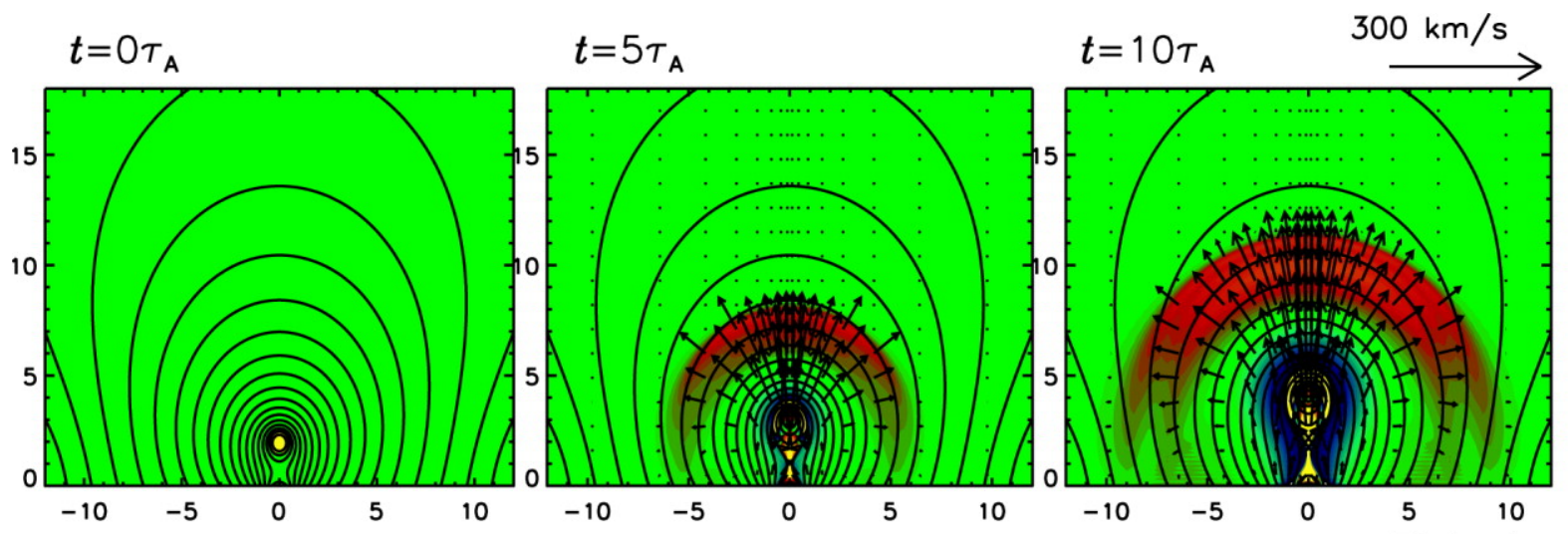}
	\includegraphics[width=0.95\textwidth,clip=,trim=0mm 0mm 0mm 0mm]{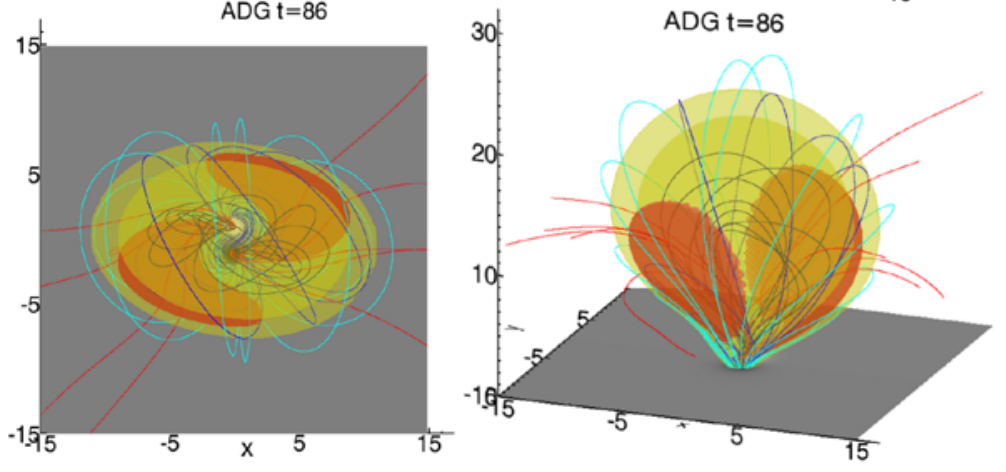}
	\includegraphics[width=0.95\textwidth,clip=,trim=0mm 0mm 0mm 0mm]{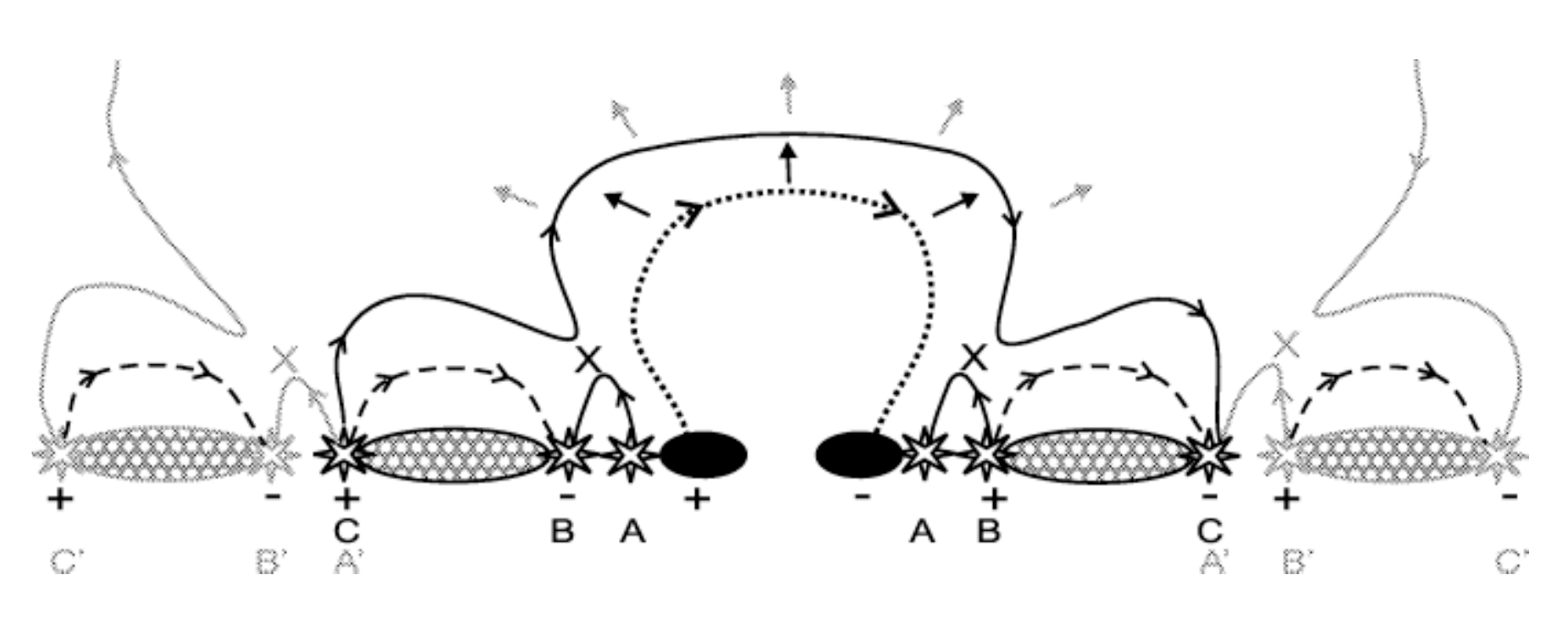}
    }
\caption{\bfref{Graphical representations of the pseudo-wave theories/models: field-line stretching model \citep[top row, 
		 modified from Figure~1 of][]{Chen:2002}; current-shell model \citep[middle row, modified from Figure~2 of][]{Delannee:2008}; 
         continuous reconnection \citep[bottom row, taken from Figure~4 of][]{Attrill:2007}.}}
\label{fig:pseudo_wave}
\end{figure*}

The field-line stretching model was originally proposed by \citet{Chen:2002} to reconcile observations indicating 
no clear relationship between solar flares and ``EIT waves'' \citep[\eg][]{Delannee:1999} and simulations of 
``EIT waves'' using unrealistic plasma-$\beta$ values \citep[\cf][]{Wu:2001}. \bfref{In order to} overcome these issues, 
\citet{Chen:2002} performed a 2D simulation using the time-dependent compressible resistive MHD equations solved 
using a multi-step implicit scheme \bfref{\citep{Hu:1989,Chen:2000}}.
This was expanded on by \citet{Chen:2005}, 
where two different field configurations were used to study the effect of neighbouring active regions. The model 
of \citet{Chen:2002} and \citet{Chen:2005} predicts two propagating features in a CME eruption: a fast-mode 
shock wave (termed ``coronal Moreton wave''), and a slower density perturbation resulting from the stretching of 
field lines overlying the erupting flux rope (termed ``EIT wave''). The fast-mode shock wave is covered in 
Section~\ref{ss:mhdwave}, while the slower density perturbation is discussed here.

With an erupting CME driving the stretching of field lines \bfref{(Figure~\ref{fig:pseudo_wave}, top row)},
the lateral velocity of the density perturbation is constrained by the geometry of 
the overlying magnetic field. Under the assumption of semicircular overlying field 
 \citep[as adopted 
by][]{Chen:2002}, the density perturbation should have a lateral velocity 
$\approx$\,1/3 that of the associated fast-mode wave. However, it is noted that this should be considered 
 an upper 
limit because the \bfref{lateral} velocity will be smaller 
 when field lines are radially elongated rather than being 
semicircular. The time taken for information on the stretching of the overlying field to reach the low corona increases 
as successively higher field lines are perturbed. Outside the source active region, this yields decreasing lateral 
velocity (\ie\ $a<0$) for the low-corona density perturbation (\ie\ at the foot-points of the overlying field).

The stretching of overlying magnetic-field lines manifests itself as an increase in magnetic-field strength, 
with the density increasing due to the frozen-in effect. Numerical simulations performed by \citet{Chen:2005} 
also suggest a weak temperature increase in the density perturbation. \bfref{The successive outward stretching of field lines has a component in all directions (\ie\ not only vertically), resulting in a broadening of the pulse in time \citep{Chen:2002b}.}

\citet{Chen:2005} noted that the dimming region within the CME bubble (characterised by strong upflows) should 
define the area enclosed by the density perturbation. This model also predicts a dome-like structure for the 
density perturbation, constrained by the height of the driving CME \citep{Chen:2009}. Rotation of the density 
perturbation is not present in 2D simulations, but it is expected in 3D when the background magnetic field is 
sheared. The density perturbation is not expected to show reflection or refraction at CH boundaries, instead 
possibly resulting in long-lived stationary brightenings at CH edges.

The stretching of magnetic field lines does not produce a shock and hence no co-spatial \rtii\ radio burst is 
expected. In addition, the density perturbation is not expected to be energetically capable of producing a 
 Moreton--Ramsey wave. However, the coronal fast-mode shock 
 also produced in this model 
\bfref{was proposed to be able to produce both 
 of these phenomena 
 if it is} of sufficient amplitude.

\subsection{Current-Shell Model}\label{ss:Delannee}

The current-shell model proposed by \citet{Delannee:2007} aimed to explain the non-isotropic nature of 
 ``EIT waves'' that they suggested was incompatible with an MHD fast-mode wave. 
 The model also 
aimed to explain previous observations of long-duration stationary bright features associated with ``EIT waves'' 
\citep{Delannee:2000}. In all three events studied by \citet{Delannee:2000}, part of the observed bright feature 
propagates while another part remains stationary from frame to frame, a behaviour \bfref{that those 
authors had suggested to be} inconsistent with the interpretation of the disturbance as a wave.

Instead, \citet{Delannee:2007} proposed that as the coronal magnetic field 
 opens during the eruption of a 
CME, electric currents should form due to sudden jumps in magnetic field connectivity. \bfref{As 
shown in the middle row of Figure~\ref{fig:pseudo_wave},} this produces a large-scale current shell that separates 
 an 
 \bfref{erupting flux rope} from the surrounding coronal 
 field. The current shell 
 dissipates 
its energy 
 via Joule heating, resulting in an EUV brightening at these locations. As a result, the 
kinematics of the wave will be consistent with the lateral expansion of the erupting CME bubble (\ie\ \vcmeperp\ 
and \acmeperp). Similarly, both the height and area of the pulse will 
 be related to the height and area of 
the CME bubble. The measured width of the pulse will strongly depend on the lateral velocity of the CME and the 
plasma cooling time. However, the 3D nature of 
 a 
 current shell means 
 there will be an additional 
contribution as the emission from several heights is combined along the 
 line of sight. The temporal 
evolution of all 
 of these components determines whether or not the pulse is observed to broaden in time.

Although an increase in density is not required to produce the Joule heating (and hence increased temperature), 
the expanding nature of the CME leads to a density enhancement due to compression at the leading edge of the CME 
bubble. The physical processes involved also suggest that it is unlikely to exhibit any clear variation in 
magnetic-field strength associated with the observed disturbance.

Rotation of the propagating disturbance was seen in simulations by \citet{Delannee:2008}, but no reflection or 
refraction is expected at CH boundaries. The combination of the current shell with the open field topology of CHs 
leads to deflection of the CME bubble or sustained reconnection at the CH boundary for unfavourable or favourable 
magnetic-field orientation, respectively. Neither of these scenarios result in transmission through the CH, while 
the latter can produce a long-lived intensity enhancement (\ie\ stationary front) at the boundary.

As with field-line stretching in Section~\ref{ss:2wave}, Joule heating itself will not produce a \rtii\ burst or 
a Moreton--Ramsey wave, as no shock is formed.

\subsection{Continuous Reconnection}\label{ss:attrill}

The ``EIT wave'' concept of continuous reconnection was originally proposed by \citet{Attrill:2007} to explain 
two of their observed properties -- the rotation of several ``EIT waves'' with propagation, and the appearance 
of remote coronal dimming regions associated with erupting CMEs. Two events from 7 April 1997 and 12 May 1997 were
chosen to illustrate these effects, as both were very clear ``EIT waves'' that appeared to rotate with propagation 
\citep[an observation originally noted by][]{Podlad:2005}. It was found that the direction of rotation exhibited by 
the ``EIT waves'' was defined by the helicity of the erupting active region. Although it had been suggested 
that ``EIT waves'' could be blast waves driven by the energy release in a solar flare, \citet{Attrill:2007} suggested 
that the observed rotation was incompatible with that interpretation. Instead, they proposed that ``EIT waves'' were 
due to reconnection between the magnetic structure of the erupting CME and the surrounding coronal field.

In the continuous reconnection model, 
 an 
 ``EIT wave'' brightening is the result of systematic reconnection between the 
erupting CME flux rope and surrounding (favourably oriented) loops and open field lines in the quiet Sun. As these field 
lines reconnect, a steady stream of particles are accelerated at the reconnection sites toward 
 the chromosphere (as in 
 solar flares, but on 
 much smaller and weaker scales). A series of brightenings will be created via chromospheric 
evaporation as these particles impact the dense lower atmosphere, with 
 brightenings observed as 
 an 
 ``EIT wave'' 
\bfref{(Figure~\ref{fig:pseudo_wave}, bottom row)}. As a result, this should exhibit 
signatures consistent with those observed during a flare (\ie\ \bfref{strongly} increased temperatures and densities due to upflowing 
hot material), albeit on a smaller scale. In addition, there may be minor variations in the magnetic-field strength.

The reconnection process between the erupting CME and the adjacent small-scale coronal loops means that the kinematics 
of the pulse should be defined by the lateral motion of the erupting CME (\ie\ \vcmeperp\ and \acmeperp). The combination 
of this lateral motion and the plasma cooling time scale [$t_{\mathrm{cooling}}$] define the apparent width of the pulse 
and its temporal variation. In addition, the area bounded by the pulse is defined by the lateral extent of the CME, with 
the pulse being formed low in the solar atmosphere at heights comparable to quiet-Sun loops. As with the current-shell 
model, the ``EIT wave'' produced by continuous reconnection is not expected to exhibit reflection or refraction at CH 
boundaries, but it can result in long-lived stationary bright fronts due to interchange reconnection 
\citep[\eg][]{Attrill:2006}.

Although the low-coronal nature of the reconnection process suggests that it may be possible for Moreton--Ramsey waves to 
be produced, this will depend upon the amount of energy released during reconnection. However, there will be no signature 
of a \rtii\ radio burst as no shock is formed in this approach.

\section{\bfref{Data} Analysis \bfref{and Modeling}}\label{s:analysis}

As summarised in Table~\ref{tbl:theories}, the 
theories proposed to explain 
 ``EIT waves'' 
 make separate predictions for physical properties, each of which may be measured and used to confirm the interpretation. 
However, the techniques used to observe and analyse ``EIT waves'' can influence the value and behaviour of the different 
properties being measured. ``EIT waves'' are traditionally observed as broad and diffuse low-intensity features that are 
difficult to identify in single intensity images, and as a result they are often identified using movies or difference images 
(where a leading image is subtracted from a following image). The temporal step used when subtracting images can affect the 
size, shape, and derived velocity of the pulse, so care must be taken when using difference images. However, the advent of 
multiple passbands for observing ``EIT waves'' and improvements in image processing and analysis techniques are providing 
an opportunity to simultaneously study multiple properties of these features, allowing a better discrimination between 
theories.

\subsection{Current \bfref{Observational and Analysis} Capabilities}\label{ss:current}

The kinematics of observed ``EIT waves'' are the easiest property to measure, and as a result they have been 
calculated since ``EIT waves'' were first observed. The multitude of proposed theories arose primarily from discrepancies 
between the observed and predicted behaviour of the pulse kinematics, and it continues to be the primary method of 
differentiating between theories. However, it is clear from Table~\ref{tbl:theories} that kinematics alone are not 
sufficient to discriminate between theories and other properties should be taken into account. It has been suggested 
that estimates of kinematics can be strongly affected by observational temporal cadence 
\citep[\eg][]{Long:2008,Byrne:2013}, although \citet{Muhr:2014} found no relationship between observing cadence and pulse 
velocity.

As previously noted, ``EIT waves'' are difficult to identify in single images, often requiring movies to enable 
identification. Extensive image processing is therefore required to allow the identification of ``EIT waves'' in single 
images. This is primarily achieved using difference or ratio images, where an image has a preceding image subtracted from 
or divided into it, respectively. These allow a feature to be identified by highlighting changes in intensity as the 
position of the feature changes between the two images. Although the temporal step between images does not affect the 
identification of the ``EIT wave'' leading edge, it can affect the observed pulse properties (\eg\ width, amplitude). 
For example, short temporal steps suppress long-term intensity variations due to motion of the background corona and 
highlight fast-moving features. Conversely, large temporal steps highlight long-term variations but become progressively less 
clear as the temporal step increases.

The optimal temporal step for estimating kinematics and properties of an ``EIT wave'' must therefore be chosen carefully. 
The temporal step must be large enough that the positions of the pulse do not overlap in the two images, as this leads to 
incorrect estimation of pulse properties such as width and peak intensity difference (or intensity ratio). Similarly, 
the temporal step must be sufficiently small that the pulse can be identified above the noise of the varying background 
corona. This approach allows an estimate to be made of the variation in pulse position and (depending on the temporal step) 
the width of the pulse with time.

After identifying 
 a pulse, 
 several 
 methods 
 can be used to study 
 it. 
 The position 
 can be identified manually in a series of images using a point-and-click approach 
 \citep[\eg][]{Narukage:2004,White:2005,Thompson:2009}, but this method 
 is user-dependent and subject to bias. 
 Recent work has tended to use automated 
 methods 
 that minimise 
user involvement and employ predefined properties of the pulse to identify and track it. Although automation makes 
them relatively self-consistent, algorithms such as \textsf{CorPITA} \citep{Long:2014} and \textsf{Solar Demon} \citep[][previously \textsf{NEMO}; 
\citet{Podlad:2005}]{Kraaikamp:2015} may not necessarily return the same values for pulse properties as those 
identified manually. \bfref{In addition, Huygens tracking was proposed to interpolate between pulse positions 
\citep{Wills-Davey:2006} but, to the best of the authors' knowledge, this approach has not been used elsewhere.}

Both \textsf{CorPITA} and \textsf{Solar Demon} measure the kinematics of the pulse along defined directions using intensity profiles, a 
technique 
 also 
 often 
 used for manual estimates of 
 pulse kinematics. 
 This 
 approach collapses 
the intensity along an arc into a 1D intensity plot \citep[\eg][]{Muhr:2011,Long:2014}, allowing the pulse to be 
identified as an increase in intensity that may be fitted and tracked using a predefined model. An alternative approach 
is a stack plot that combines 1D profiles into a 2D image \citep[\eg][]{Liu:2010,Ma:2011,Shen:2012}, allowing 
the pulse to be visually identified. Both approaches have merit: 
 1D spatial 
 profiles allow the amplitude/width/shape of 
 a 
 pulse to be tracked and studied; stack plots allow the identification of additional fronts that may result 
from 
 projection effects as 
 a 
 CME erupts. \bfref{Stack plots have also been shown to be useful when looking 
for reflection and refraction of wave pulses at CH and active-region boundaries 
\citep[\eg][]{Gopal:2009,Olmedo:2012,Kienreich:2013}, while slicing along the temporal axis yields a temporal profile from 
one location, enabling studies of pulse passage effects on the background corona}.

The volume of data and improved spatial and temporal resolution now available from current and upcoming instrumentation 
provide an opportunity for improved discrimination between different theories. It is now possible to visualise 
heating and cooling in the pulse using the subtly different temperature responses of the different SDO/AIA passbands. 
As described by \citet{Downs:2012}, the 171\,\AA, 193\,\AA, and 211\,\AA\ passbands may be combined and used to 
indicate changes in the temperature of the plasma during the passage of an ``EIT wave''. This may be used as an 
additional constraint on the interpretation of the feature.

A more detailed analysis to quantify the amount of heating or cooling requires a full estimation of the differential 
emission measure (DEM) of the plasma. Although this requires knowledge of a series of emission lines obtained from 
spectroscopy, recent advances in analysis techniques and the temperature ranges covered by the multiple EUV passbands 
of SDO/AIA are beginning to allow the calculation of DEMs using broadband images. \citet{Kozarev:2011} used DEM 
analysis of a limb eruption on 13 June 2011 to find a density increase of 12\,\%, while \citet{Vanninathan:2015} used a 
similar approach for an on-disk event on 15 February 2011. Using the regularized-inversion technique of 
\citet{Hannah:2012,Hannah:2013}, they found a density increase of 6\,--\,9\,\%, corresponding to a temperature increase 
of $\approx$\,5\,--\,6\,\% during the ``EIT wave'' passage, which was shown to be a result of adiabatic compression at the 
wave front.

Although this approach 
 provides 
 an estimate of the variation in temperature and density 
associated with the ``EIT wave'', it is restricted by the broadband nature of the instrument that sacrifices spectral 
resolution for spatial resolution. The 
 alternative 
 to this is provided by slit 
 spectrometers such as the 
\emph{Extreme-ultraviolet Imaging Spectrometer} \citep[EIS:][]{Culhane:2007} onboard the \emph{Hinode} 
 spacecraft 
\citep{Kosugi:2007}. EIS has 
 very high spectral resolution from a 1\,--\,2\arcsec-wide slit that provides an 
opportunity to identify 
 up-/down-flows associated with the passage of the pulse. However, the very small 
field-of-view of EIS and the anisotropic nature of ``EIT waves'' make it difficult to observe a pulse, and 
spectrometric observations of ``EIT waves'' remain extremely rare. Despite this, EIS has observed at least two 
 events, 
both of which have been extensively studied \citep[\cf][]{Harra:2011,Veronig:2011,Chen:2011,Long:2013}.

The ``EIT waves'' studied using DEM techniques and observed by \emph{Hinode}/EIS exhibited signatures consistent with 
a shock wave interpretation, including rapid temperature and density increases and very high velocities. However, the 
primary signatures used to indicate the presence of a shock wave are \rtii\ radio bursts 
\citep[\eg][]{Klein:1999,Mann:1995b,Mann:1995c}. Dynamic radio spectra are sufficient to identify the 
existence of a shock wave and potentially its height (inferred from a density model or estimate), but they do not 
 identify 
 the spatial location of the shock. However, this can be achieved \bfref{occasionally} using radio imaging observations from, \bfref{\eg,} 
the Nan\c{c}ay Radioheliograph. This facility was used by \citet{Pohj:2001}, \citet{Khan:2002}, 
\citet{Vrsnak:2005,Vrsnak:2006}, and \citet{Carley:2013} to identify and track \rtii\ radio emission associated with 
``EIT waves'', 
 complemented by dynamic spectra from multiple instruments to study the associated 
shock-accelerated particle emission. 
\bfref{Unfortunately, radio imaging of \rtii\ bursts and CMEs is exceptionally rare due to the lack of dedicated solar facilities around the world, while \rtii\ emission is a sufficient but not necessary condition for the existence of a shock.}


Although chosen as the primary discriminant between different theories and models of ``EIT waves'', the kinematics of 
the pulse cannot and should not be considered as conclusive proof that the feature conforms to one interpretation at 
the expense of all others. The interaction of the pulse with the surrounding corona is also of vital importance, with 
signatures of reflection and/or refraction providing an opportunity to immediately discriminate between wave and 
pseudo-wave interpretations. Any study of ``EIT waves'' should \bfref{also} include the variation of temperature and density of 
the plasma associated with the propagation of the pulse and radio observations of a shock, or lack thereof. A 
combination of current instruments provides multiple discriminants that may be used to add to the weight of evidence 
in favour of one theory or another. In addition, there may be other properties or observations beyond our current 
capability \bfref{(\eg\ changes in coronal magnetic field)} that could be used to definitively identify the physical processes involved in the development and 
propagation of an ``EIT wave''.

\subsection{Modeling and Simulations}\label{ss:modeling}

An additional tool that may be used to understand and interpret ``EIT waves'', and indeed potentially discriminate 
between theories, is modeling of the eruption and evolution of the ``EIT wave'' itself. This is not a new approach, 
with many of the original theories designed to explain the phenomenon proposed following simulations of solar 
eruptions \citep[\eg][]{Delannee:2000,Chen:2002}. The complexity and realism of the simulations and modeling vary 
widely, with both simple analytical modeling and more complex numerical modeling including 3D MHD models providing 
different insights into the processes involved.

Most of the analytical modeling has focused on characterising how the large-amplitude wave front forms as a result of 
the explosive expansion of a three-dimensional piston. This approach has been studied in detail by multiple groups, 
with \citet{Vrsnak:2000a,Vrsnak:2000b} in particular describing how this process evolves. More recent work has 
focused on the details of the formation mechanism, the time scales over which the wave front forms and the exact 
nature of the original piston \citep[\eg][]{Zic:2008,Temmer:2009,Lulic:2013}.

An alternative approach is to model the formation and evolution of the ``EIT wave'' numerically. As described by 
\citet{Vrsnak:2016}, this can be done in one of two ways: either using realistic configurations for the initial 
eruption and the background corona to study specific events 
\citep[\eg][]{Cohen:2009,Schmidt:2010,Downs:2011,Downs:2012}, or alternatively using a simplified configuration to 
understand the general processes involved \citep[\eg][]{Chen:2002,Chen:2005,Wang:2009,Hoilijoki:2013,Wang:2015}. 
Both approaches offer unique insights into the initiation and evolution of the pulse and the effects of the 
background corona.

Simulations and modeling provide a unique opportunity to examine the behaviour of ``EIT waves'' and compare the 
observations with that predicted by theory. In particular, combining simulations with the different observations 
offered by instruments such as SDO/AIA, \emph{Hinode}/EIS, and the twin viewpoints of the STEREO spacecraft offer a 
powerful new tool to examine this phenomenon. 

\subsection{Future Diagnostics}\label{ss:future}

While current instrumentation and techniques provide an unprecedented view of the Sun, particularly the low corona, 
some gaps in observations remain. \bfref{This is especially true of the coronal magnetic field, with 
Table~\ref{tbl:theories} identifying different behaviours for the theories/models (notably providing discrimination in 
both the wave and pseudo-wave categories)}. Current 
 magnetic-field observations use Stokes polarization measurements and the Zeeman effect to estimate field 
strength/orientation in the photosphere where the signal is strong. However, 
 low 
signal-to-noise in coronal emission lines means that, rather than measuring the coronal 
 field directly, 
photospheric field must be extrapolated 
 to infer the coronal field, particularly outside active 
regions as the magnetic field is relatively weak. 

Some advances have been made in 
 \bfref{measuring} the coronal magnetic field \bfref{-- \eg,} 
 the \emph{Coronal Multi-channel 
Polarimeter} \citep[CoMP:][]{Tomczyk:2008} 
 that uses 
 the \ion{Fe}{13} 10747\,\AA\ emission line to measure 
the Stokes polarization in the low 
 corona at heights of $\approx$\,1.03\,--\,1.5\,R$_{\odot}$. However, the 
signal-to-noise for CoMP is quite low and, combined with the ground-based nature of the instrument, this makes observations 
of ``EIT waves'' with CoMP \bfref{extremely challenging}. One of the aims of the forthcoming \emph{Daniel K. Inouye Solar 
Telescope} (DKIST) is to estimate the coronal magnetic field, with its 4-m primary mirror greatly improving the signal-to-noise 
ratio. As a result, it may be possible to \bfref{measure} variations in coronal magnetic-field \bfref{strength} during the 
\bfref{passage} of an ``EIT wave'', \bfref{enabling the $\Delta B$ row in Table~\ref{tbl:evidence_base} to be addressed. However, the 
small field-of-view for DKIST (\ie\ $<$\,100\arcsec) means that ``EIT wave'' observations will most likely be serendipitous.}


\section{Discussion}\label{s:disc}

The 
 range 
 of theories proposed to explain ``EIT waves'' (outlined in Section~\ref{s:theory}) and the number 
of articles and reviews devoted to this phenomenon 
 show 
 that they remain a subject of interest to the 
broader community. However, it is also clear that many of the original interpretations were affected by the 
relatively low temporal and spatial resolutions of SOHO/EIT. 
 Multi-point observations from the STEREO spacecraft 
 provide better insight into the relation between 
 ``EIT waves'' and 
 CMEs, while the improved temporal/spatial 
capabilities of SDO/AIA could supply sufficient evidence 
 of 
 the 
 physical processes 
 at work.

Table~\ref{tbl:theories} was constructed to indicate the different properties and behaviours that may be used to 
discriminate between the different theories. How each of these predictions compares to observations is presented in 
Table~\ref{tbl:evidence_base} with symbols indicating that observations and predictions are in agreement (\gtick), 
observations are not inconsistent with predictions ($\sim$), measurements have not been or cannot yet be made (--), 
and observations do not match predictions (\rcross). The properties given in Tables~\ref{tbl:theories} and 
\ref{tbl:evidence_base} can be grouped into five main headings, each of which is discussed in more detail below: 
kinematic properties, physical properties, geometric properties, spatio-temporal properties, and associated phenomena.

\ctable[
caption = {Observational support for model/theory predictions: agreement (\gtick); not inconsistent ($\sim$); no 
measurements available (--); no agreement (\rcross)},
label = tbl:evidence_base,
center,
sideways
]{lcccccc}{
\tnote[$^\mathrm{a}$]{describing only the \bfref{slower component of the two-wave scenario (\ie\ the density perturbation component)}}
}{
\FL
Pulse                        & \multicolumn{3}{c}{Wave theories}         & \multicolumn{3}{c}{Pseudo-wave theories/models}\NN
physical                     & \multicolumn{2}{c}{Fast-mode} & Slow-mode & Field-line                      & Current & Continuous\NN
property                     & Small amp.  & Large amp.      & soliton   & stretching\tmark[$^\mathrm{a}$] & Shell   & reconnection\NN
                             & linear wave & wave/shock      &           &                                 &         & 
\ML
Phase velocity, $v$  & \gtick      & \gtick          & $\sim$    & $\sim$                          & \rcross & \rcross\NN
Acceleration, $a$    & \gtick      & \gtick          & \gtick    & \rcross                         & \rcross & \rcross\NN
Broadening           & \gtick      & \gtick          & \gtick    & \gtick                          & --      & --\NN
$\Delta B$           & --          & --              & --        & --                              & --      & --\NN
$\Delta T$           & \gtick      & \gtick          & \gtick    & \gtick                          & \gtick  & \rcross\NN
$\Delta \dens$       & \gtick      & \gtick          & \gtick    & \gtick                          & \gtick  & \rcross\NN
Height               & \gtick      & \gtick          & \gtick    & \gtick                          & \rcross & \rcross\NN
Area bounded         & \gtick      & \gtick          & \gtick    & \rcross                         & \rcross & \rcross\NN
Rotation             & $\sim$      & $\sim$          & $\sim$    & \gtick                          & \gtick  & \gtick\NN
Reflection           & \gtick      & \gtick          & \gtick    & \rcross                         & \rcross & \rcross\NN
Refraction           & \gtick      & \gtick          & \gtick    & \rcross                         & \rcross & \rcross\NN
\bfref{Transmission} & \gtick      & \gtick          & \gtick    & \rcross                         & \rcross & \rcross\NN
Stationary fronts    & \gtick      & \gtick          & \rcross   & \gtick                          & \gtick  & \gtick\NN
Co-spatial \rtii     & \rcross     & \gtick          & \rcross   & \rcross                         & \rcross & \rcross\NN
Moreton wave         & \rcross     & \gtick          & \rcross   & \rcross                         & \rcross & \gtick
\LL
}

\subsection{Kinematic Properties}

As the easiest property to identify for a global pulse, the velocity [$v$] predicted for each theory is well-defined 
in Table~\ref{tbl:theories}. As fast-mode waves, the small- and large- amplitude waves should propagate at or somewhat 
above the fast-mode wave velocity, respectively. Statistical studies of ``EIT waves'' \bfref{using SOHO/EIT 
and STEREO/EUVI found average velocities of 200\,--\,500\,\kms\ \citep[\cf][]{Klassen:2000,Thompson:2009,Muhr:2014}, consistent with fast-mode waves. 
Although recent studies using SDO/AIA report higher average velocities of 600\,--\,730\,\kms\ 
\citep{Nitta:2013,Liu:2014}, it should be noted that these consider the maximum velocity in any direction 
rather than directional averages like earlier works.} 
 While some studies indicate 
 higher-intensity pulses have greater velocities \citep{Muhr:2014}, pulse intensity 
may be affected by the background corona \citep{Nitta:2013} making comparison between pulse amplitude and velocity 
difficult. 
 Hence, 
 observations are not inconsistent with the MHD slow-mode soliton. 
\bfref{Existing studies of lateral CME expansion \citep[\eg][]{Patsourakos:2010} clearly show the formation and decoupling of the 
``EIT wave'' from the CME driver, behaviour inconsistent with the predictions made by the MHD slow-mode soliton and pseudo-wave models.}

All of the wave theories predict pulse accelerations [$a$] that are either less than zero or equal to zero, matching 
results found by multiple authors 
\citep[\eg][]{Warmuth:2004a,Long:2011b,Zheng:2012a,Nitta:2013}. 
\bfref{Distinct correlations are found between ``EIT wave'' initial velocities and acceleration -- faster events show stronger 
deceleration, while those near quiet-Sun coronal fast-mode speeds have no significant deceleration \citep{Warmuth:2011,Muhr:2014} -- a relation expected for large-amplitude 
 fast-mode waves. In contrast}, 
the field-line stretching, current-shell, and continuous reconnection models predict \bfref{bright-front} acceleration 
equal to the lateral acceleration of the expanding CME. 
\bfref{The observed process of ``EIT waves'' decoupling from the CME, as discussed above, requires a difference in lateral acceleration 
 that is inconsistent with the pseudo-wave models}.

\subsection{Physical Properties}

Broadening (or not) of the pulse with propagation is well-defined for the wave and 
field-line stretching models. MHD slow-mode solitons and small-amplitude 
 fast-mode waves predict a pulse with 
minimal or no broadening, while the large-amplitude fast-mode wave and field-line stretching models 
 predict 
clear pulse broadening with propagation. This behaviour has been reported 
 multiple 
 times 
\citep[\eg][]{Warmuth:2004b,Long:2011a,Muhr:2011}, supporting the wave and field-line stretching models. Although 
they do not require a pulse to broaden with propagation, such behaviour is not inconsistent with the current-shell 
and continuous reconnection models. For these, broadening 
 depends 
 on 
 the lateral acceleration of the 
CME (current-shell model) or the lateral velocity of the CME and cooling time of the plasma (continuous reconnection 
model). 

Each theory predicts that ``EIT waves'' will have an increase in temperature and density, although the mechanisms 
producing these 
 differ. The wave, field-line stretching, and current-shell models yield density 
increases from compression, while for continuous reconnection it is upflowing chromospheric plasma. Compression is clearly observed in ``EIT waves'', supporting the wave, field-line 
stretching, and current-shell models \citep[\cf][]{Kozarev:2011,Ma:2011}. 
 Spectroscopic observations show downflows \bfref{of up to 20\,\kms\ at the ``EIT wave'' front 
\citep{Harra:2011,Veronig:2011}, indicative of downward plasma motion related to 
 compression and inconsistent with continuous reconnection.} 

Recently \citet{Schrijver:2011} and \citet{Vanninathan:2015} have shown that temperature enhancements in ``EIT waves'' 
are due to adiabatic heating, consistent with the wave, field-line stretching, and \bfref{partially also the 
current-shell model. This is inconsistent with the continuous reconnection model that predicts an increase 
 due to 
non-adiabatic processes (\ie\ low-energy magnetic reconnection). The temperature increase reported by \citet{Vanninathan:2015} 
was highest at the peak and frontal part of the ``EIT wave". 
 While such behaviour is expected for a compressive 
fast-mode wave and the field-line stretching model, the current-shell model predicts the largest increase in the rear 
part of the ``EIT wave". The temperature enhancement is an adiabatic process in the fast-mode wave and field-line 
stretching models, such that the largest temperature increase occurs with the largest density increase (\ie\ cospatial 
with the EUV wave). In the current-shell model, the ``EIT wave" pulse is explained by Joule heating in the current 
shell that builds up at the separation layer between the erupting flux rope and the surrounding field. Thus, the 
temperature increase is expected to be highest at this CME-wave interface (\ie\ in the rear portion of the EUV wave).}

Although each of the models makes a distinct prediction for the variation in magnetic field across the ``EIT wave'' 
pulse, this property cannot be measured using current observational techniques. As noted in Section~\ref{ss:future}, 
measuring the magnetic field in the corona is hampered by the very low signal-to-noise of magnetically-sensitive 
coronal lines. Detection of a small perturbation in the magnetic field during the passage of a fast, diffuse wave is therefore 
beyond the capabilities of current instrumentation, and thus these predictions cannot currently be tested.

\subsection{Geometric Properties}

The wave theories all suggest that the height over which the wave is observed should vary as a function of the 
magnetic field and density of the background corona, while the area bounded by the pulse should exceed the area 
bounded by the associated CME. In contrast, the pseudo-wave theories predict an ``EIT wave'' with a height 
defined by the CME leading loop (field-line stretching model), at fairly constant heights of $\approx$\,280 or 407\,Mm 
\citep[current-shell model; values from][]{Delannee:2008}, or at heights $<$\,10\,Mm \citep[continuous 
reconnection model; value from][]{Patsourakos:2009b}. In addition, the area bounded by the ``EIT wave'' 
should equal the area of the CME in all pseudo-wave cases.

The visibility of a 3D wave perturbation in a gravitationally stratified corona is expected to be weighted by 
density (which controls emissivity), with the \bfref{density} scale height being $\approx$\,70\,--\,90\,Mm for typical 
quiet-Sun temperatures of $\approx$\,1.0\,--\,1.7\,MK. These heights are roughly consistent with the observed 
heights of ``EIT waves'' measured using intensity-based diagnostics 
\citep[\cf][]{Patsourakos:2009a,Patsourakos:2009b,Kienreich:2009} 
 that 
 are significantly different from the 
predictions of the current shell and continuous reconnection models. Observations of dome-shaped ``EIT waves'' 
\citep[\eg][]{Veronig:2010} and dome-shaped fronts connecting to features in the extended corona 
\citep[\eg][]{Cheng:2012,Kwon:2014} are 
 consistent with the wave 
 expectations of a 3D perturbation 
being present at larger heights. In addition, \bfref{after decoupling,} the area bounded by an ``EIT wave'' 
 exceeds 
that of the associated CME \citep[\eg][]{Patsourakos:2009a} and 
 coronal dimming region 
\citep[\eg][]{Veronig:2010}, supporting the 
 wave-theory predictions. 

\subsection{Spatio-Temporal Properties}

Spatio-temporal properties of ``EIT waves'' feature prominently in predictions of the different theories, 
unsurprising given that they are moving features. The most obvious are reflection, refraction, and transmission 
of the pulse when faced with \bfref{an active region or CH boundary}. These properties provide the starkest contrasts between the wave and 
pseudo-wave 
 branches, 
with 
 wave theories all predicting 
 ``EIT waves'' can show reflection and 
refraction 
 and 
 are expected to transmit, 
 while 
no pseudo-wave theories predict 
 \bfref{these behaviours. However, 
observations provide clear evidence of reflection \citep[\eg][]{Gopal:2009,Kumar:2013}, refraction 
\citep[\eg][]{Ofman:2002,Shen:2012b}, and transmission \citep[\eg][]{Olmedo:2012,Shen:2013}.}

Interaction between ``EIT waves'' and CH boundaries leads to observations of stationary bright fronts at the 
edges of CHs. \bfref{This was initially attributed to being due to reconnection between 
the erupting CME and magnetic field in the CH \citep[\eg][]{Delannee:2000,Delannee:2007,Attrill:2007}. 
Recently, \citet{Kwon:2013} reported on observations of stationary fronts at CH boundaries in the low 
corona while the associated ``EIT wave'' continued to propagate in the higher corona through magnetic 
streamers. This behaviour is supported by} simulations of wave pulses interacting with sudden changes in 
density and magnetic field \bfref{that} can produce stationary bright features at the interface region 
(B. Vr\v{s}nak private communication, 2016). Although MHD slow-mode solitons by definition should not 
show any variation due to a sudden change in density and/or magnetic field, it may be possible for a small- or 
large- amplitude fast-mode wave to produce a short-lived stationary bright feature.

Observations that were first reported by \citet{Podlad:2005} and subsequently by \citet{Attrill:2007,Attrill:2014} 
have suggested that some ``EIT waves'' exhibit rotation during their propagation. This is 
most consistent with the eruption of a rotating CME, and it matches predictions made by the field-line stretching, 
current-shell, and continuous-reconnection models. However, it could also be explained by the rotation of an 
erupting elliptical CME that drives a wave pulse that subsequently propagates freely. As a result, rotation 
(or lack thereof) is not inconsistent with the wave interpretations.

\subsection{Associated Phenomena}

As well as the CME associated with the propagating ``EIT wave'', the proposed theories must account for co-spatial 
\rtii\ radio emission and Moreton--Ramsey waves. The only theory that can definitely produce \rtii\ radio emission 
that is co-spatial with an ``EIT wave'' is that of a large-amplitude wave/shock, as the generation of a \rtii\ 
burst requires the formation of a shock. This has been confirmed by multiple observations 
\citep[\eg][]{Pohj:2001,Khan:2002,Vrsnak:2005,Vrsnak:2006,Carley:2013} providing evidence of a co-spatial \rtii\ 
radio burst that track the propagating ``EIT wave''.

Despite observations of Moreton--Ramsey waves being rare, they always have associated and co-spatial ``EIT waves'' 
with kinematics consistent between the two phenomena 
\citep[\eg][]{Warmuth:2001,Veronig:2006,Muhr:2010,Asai:2012}. The large-amplitude wave/shock 
interpretation can generate a Moreton--Ramsey wave, as it can have sufficient energy to compress the upper 
chromosphere and produce the required ``down-up'' \bfref{plasma motions} observed in H$\alpha$ spectra
 \citep[\cf][]{Dodson:1964,Vrsnak:2016}. 
 \bfref{Simultaneous H$\alpha$, EUV, and radio imaging observations have confirmed the co-occurrence of these 
 phenomena \citep[\eg][]{Thompson:2000,White:2005}.}
Although the field-line stretching and current shell that occur during the eruption of a CME cannot produce a 
Moreton--Ramsey wave, the process of continuous reconnection could produce small-scale chromospheric brightenings 
that might be interpreted as a Moreton--Ramsey wave.

\section{Conclusions}\label{s:conc}

In this article we have identified 15 \bfref{fundamental} ``EIT wave'' properties that may be used to discriminate 
between the different theories proposed to explain them. These are outlined in 
Table~\ref{tbl:theories} and include kinematic, physical, geometric, and spatio-temporal properties of ``EIT waves'' 
and their associated phenomena. The properties have been characterised using the original articles that proposed the 
theories and a detailed investigation of the 
 physics 
 underpinning 
 each interpretation. \bfref{Although 
this list may not be exhaustive, we believe this table} provides all of the necessary information to discriminate 
between interpretations and should be used when determining the nature of the global wave pulse being studied. 

The techniques employed to measure these properties can also have significant impact on their accuracy. 
Different analysis techniques are optimised for studying different properties and care is needed to ensure 
the most appropriate are used. Also, it may not be possible to measure some properties using the techniques
available. For example, coronal magnetic-field strength is currently very 
difficult to measure outside active regions, forcing assumptions to be made about the pulse in order to estimate 
the magnetic field strength via seismological techniques \citep[\eg][]{Mann:1999,Warmuth:2005,Long:2013}. Although 
it is important for understanding the structure of the quiet solar corona and how that can affect the 
directionality of eruptions \citep[\cf][]{Mostl:2015}, it precludes using observed changes in field 
strength to diagnose the nature of the pulse. Also, while trends in temperature and density can be 
estimated via imaging in multiple EUV passbands (\eg\ SDO/AIA), precise measurements require spectroscopy, 
which is hampered by small fields-of-view available to slit instruments.

Despite these limitations, Table~\ref{tbl:evidence_base} shows how the predictions of the proposed theories 
stand up to existing observations. Although it is not currently possible to test some of the predictions, or 
some properties have not yet been measured, the vast majority have been tested. 
 Given the 
content of Table~\ref{tbl:evidence_base}, most of the authors conclude that 
 propagating ``EIT 
wave'' pulses are most consistent with the fast-mode large-amplitude wave/shock interpretation, while 
\bfref{P.F.~Chen insists that two types of EUV wave should be discriminated and only the faster component can be described as a fast-mode wave \citep{Chen:2016}}.
While there may be stretching of field lines during the CME eruption and/or formation of a current 
shell with associated Joule heating and/or reconnection between the erupting CME and the surrounding corona, 
all of these processes would apply to the CME bubble rather than the pulse that is occasionally 
observed to propagate ahead of it.

%
 \begin{acks}
The authors thank the International Space Science Institute (Bern, Switzerland) for supporting the 
International Working Team on ``The Nature of Coronal Bright Fronts'' led by D.M.~Long and D.S.~Bloomfield. \bfref{The 
authors also wish to thank the 
 referee for their comments that helped to 
 improve 
the article.} D.M.~Long and A.M.~Veronig received funding from the European Community's Seventh Framework Programme under 
grant agreement No.~284461 (eHEROES project), while D.M.~Long is currently a Leverhulme Early-Career Fellow. 
D.S.~Bloomfield received funding from the ESA PRODEX Programme and the European Union's Horizon 2020 research and 
innovation programme under grant agreement No.~640216 (FLARECAST project). P.F.~Chen was supported by the 
grants NSFC 11533005 and 11025314. C.~Downs was supported by the NASA Living With a Star Program (NNX14AJ49G). 
R.-Y.~Kwon acknowledges support from the Office of Naval Research and George Mason University. 
K.~Vanninathan and A.M.~Veronig acknowledge funding by the Austrian Space Applications Programme of the Austrian Research 
Promotion Agency FFG (ASAP-11 4900217) and the Austrian Science Fund FWF (P24092-N16). B.~Vr\v{s}nak and T.~\v{Z}ic received 
funding from the Croatian Science Foundation under project 6212 ``Solar and Stellar Variability''. A.~Warmuth 
was supported by the German Space Agency DLR under grant No. 50 QL 0001. A.~Vourlidas acknowledges support from 
NASA contract S-136362-Y to NRL and APL internal funds. The top and bottom panels of Figure~2 are reproduced with 
permission of the AAS. 
 \end{acks}

\section*{Disclosure of Potential Conflicts of Interest}
The authors declare that they have no conflicts of interest.

%
 \bibliographystyle{spr-mp-sola}
 \bibliography{issi_bib}  

\end{article} 
\end{document}